# Optimized Composition: Generating Efficient Code for Heterogeneous Systems from Multi-Variant Components, Skeletons and Containers

Christoph Kessler
*IDA, Linköping University*
*S-581 83 Linköping, Sweden*
christoph.kessler@liu.se

Usman Dastgeer
*IDA, Linköping University*
*S-581 83 Linköping, Sweden*
usman.dastgeer@liu.se

Lu Li
*IDA, Linköping University*
*S-581 83 Linköping, Sweden*
lu.li@liu.se

*Abstract*—In this survey paper, we review recent work on frameworks for the high-level, portable programming of heterogeneous multi-/manycore systems (especially, GPU-based systems) using high-level constructs such as annotated user-level software components, skeletons (i.e., predefined generic components) and containers, and discuss the optimization problems that need to be considered in selecting among multiple implementation variants, generating code and providing runtime support for efficient execution on such systems.

## I. Introduction

Heterogeneous manycore systems offer a remarkable potential for energy-efficient computing and high performance; in particular, GPU-based systems have become very popular. However, efficient programming of such systems is difficult as they typically expose a low-level programming interface and require special knowledge about the target architecture. Even when using OpenCL, a portable programming framework with low abstraction level, code optimizations are still necessary for efficient execution on a specific target system architecture and configuration.

GPGPU techniques have made it possible to port general computations to GPUs; however, different types of processors require different programming models and have different performance characteristics. This motivates equipping a computation task with multiple implementation variants which can execute on different types of processors, possibly using different algorithms and/or optimization settings. Such a multi-variant task, e.g. a call to a software component with multiple implementation variants, can now be scheduled on an arbitrary supported processing unit, which leads to a better use of heterogeneity and load balancing. The existence of multiple variants not only improves portability across configuration-diverse heterogeneous systems, but also exposes optimization opportunities by choosing the fastest variant.

Often, determining the best implementation variant is a complex problem, because the decision depends on both the invocation context's properties (such as problem size, data distribution etc.) and the hardware architecture (such as cache hierarchy, GPU architecture etc.). Context dependence affects the tasks of selecting the best implementation variant for each call, as well as scheduling and resource allocation for them and the optimization of operand data transfers between different memory units. We refer to this combined problem of code generation and optimization as *optimized composition*.

In this survey paper we give a review of component-based programming frameworks for heterogeneous multi-/manycore (esp., GPU-based) systems and of the specific problems and techniques for generating efficient code. In particular, we will elaborate on annotated multi-variant components and their composition (Section II), multi-variant skeletons (Section III), and smart containers for runtime memory management and communication optimization (Section IV). As case studies we review concrete frameworks developed in the EU FP7 projects PEPPHER and EXCESS, and discuss selected related work at the end of each section. Section V considers global coordination and composition across multiple calls. Section VI concludes the paper.

## II. Composing Annotated Components

We refer to a *software component* as an application-level building block encapsulating one or several implementation variants that all adhere to the same calling interface and that are (by the grouping as a component) declared to be computationally equivalent. Different implementation variants of a component might target different types of execution units (e.g. CPU, GPU, ...), be written in different programming models, use different algorithms and/or compiler optimizations etc., all of which will result in different resource requirements and performance. For instance, a sorting component may contain several implementations using different sorting algorithms including some GPU-specific variants such as GPU-quicksort [7] or GPU-samplesort [28]. The possibility to switch for each call to a component between its multiple implementations depending on the software (call parameters) and hardware (available resources) context opens an important optimization potential.

In order to make component implementation switching easier, the PEPPHER framework [6] provides a toolchain to raise the programming level of abstraction and improve performance portability. Since the information needed for implementation selection can not be fully extracted or inferred from current programming languages, extra information needs to be provided in some form, such as XML annotations, pragmas etc. The PEPPHER composition tool [16] adopts the pure XML way for the reason that it is non-invasive to the existing components either in source code or binary form, while the Global Composition Framework [13] and the PEPPHER transformation tool [5] allow pragma-based annotations of component calls in the source code to convey additional information for composition.

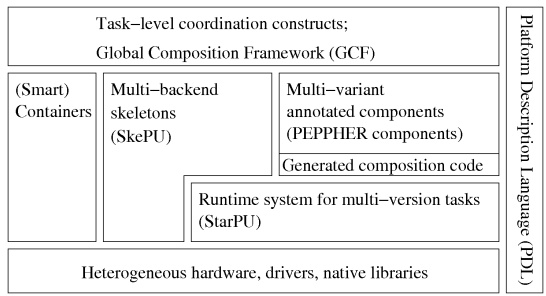

Figure 1. Software stack of PEPPHER technologies

Composition can be performed at different stages, at design time, deployment (compile) time, or run time. Design-time composition is the traditional way but lacks flexibility in a performance optimization context. Compile-time composition is more flexible since extra knowledge about the target architecture can be available. Run-time composition is the most flexible way since at this stage all information necessary for the composition choice is available, but the drawback is the run-time overhead.

The *PEPPHER component model* [6] defines how to annotate components with implementation variants written in various C/C++ programming models such as OpenMP, CUDA or OpenCL, with XML metadata required for their deployment and optimized composition.

The *PEPPHER composition tool* [16] is a compiler-like, static application build tool which generates, from the metadata, glue code containing implementation selection logic and translates component calls into task form, as required by the PEPPHER run-time system, StarPU. See also Fig. 1.

The *StarPU* runtime system for heterogeneous multicore systems [3] provides a low-level API for specifying multi-variant tasks and supports dynamic platform and implementation selection, automatic data movement management and dynamic scheduling. It uses online sampling and tuning techniques with regression-based performance models to improve its selection over time as more tasks are executed.

The selection by the composition tool can be performed statically based on programmer hints, or dynamically either by the prediction based on the off-line training data (as considered further in this work), or delegate the choice further to the PEPPHER run-time system.

Operands of component calls are passed either as native C/C++ data types or in so-called smart containers such as *Vector* and *Matrix* (see Section IV), which internally handle memory management, synchronization and communication. While PEPPHER components are generally stateless, their operands do have state, which is encapsulated in the smart containers.

*Conditional composition* [14] is a portable annotation method that allows the user to extend and guide the composition mechanism by specifying constraints on selectability for individual implementation variants, e.g. constraints on the platform (architecture and system software, e.g. installed libraries) or on the run-time context, e.g. on call parameter values or runtime system state. The PEPPHER composition tool implements conditional composition by reading in a formal model of the target hardware (in the XML-based platform description language PDL [33]) and reifying runtime data so they can be accessed in the selectability constraints. Conditional composition also provides an escape mechanism for a flexible extension of the component model to address specific cases without having to introduce new annotation syntax.

As an alternative to on-line performance modeling and tuning in the runtime system, the PEPPHER composition tool also supports off-line tuning, precomputing a tree-based dispatch structure for fast runtime lookup of the expected fastest variant from a number of sample executions. Adaptive sampling [29] is a method to trade a small imprecision in the resulting predictor for a significant reduction of the number and cost of off-line sample executions and of the remaining runtime selection overhead.

*Related work*

Kessler and Löwe [26], [27] provide a framework for optimized composition of parallel components that are non-invasively annotated by user-specified meta-data such as time prediction functions and independent call markup. Prediction functions might come from multiple sources such as microbenchmarking, measuring, direct prediction or hybrid prediction. In contrast to PEPPHER, recursive components are supported, which makes optimized composition even more powerful because its performance gains multiply up along the paths in the recursion tree. For small context parameter values (e.g. problem sizes), the best candidate is precalculated off-line by a dynamic programming algorithm and stored in a lookup table that can be compressed by machine learning techniques [10]; for larger ones, simple extrapolation is used at runtime. Then the implementation selection code based on those predictions will be generated, injected and executed dynamically.

Other systems are more intrusive so that code in legacy programming models cannot be directly reused within components. For instance, *PetaBricks* [2] suggests to expose the implementation selection choices directly in a new programming language with compiler support. PetaBricks provides an off-line tuner that samples execution time for various input sizes; the composition choice is made off-line based on the tuning results.

The Merge framework [30] targets a variety of heterogeneous architectures, while focusing on MapReduce [18] as a unified, high-level programming model. Merge provides a predicate mechanism that is used to specify the target platform and restrictions on the execution context for each implementation variant; this is somewhat similar to the conditional composition capability in the PEPPHER composition tool, which however can address also other computation patterns beside MapReduce. Merge selects between implementations by issuing work dynamically to idle processors (dynamic load balancing) while favoring specialized accelerator implementations over sequential CPU implementations. In contrast, PEPPHER's selection mechanism is based on performance models, which is more flexible.

Wang et al. [36] propose EXOCHI, an integrated programming environment for heterogeneous architectures that offers a POSIX shared virtual memory programming abstraction. The programming model is an extension of C++ and OpenMP where (domain-specific) portions of code, to be executed on a specific accelerator unit, are embedded/inlined inside OpenMP parallel blocks. The EXOCHI compiler injects calls to the runtime system and then generates a single fat binary consisting of executable code sections corresponding to the different accelerator-specific ISAs. During execution, the runtime system can spread parallel computation across the heterogeneous cores.

*Elastic computing* [37] provides a framework which enables the user to transparently select and utilize different kinds of computational resources by a library of elastic functions (implementation variants). The framework also contains an empirical autotuner and employs linear regression for performance prediction.

Grewe and O'Boyle [23] suggest a static method for partitioning dataparallel OpenCL computations to decide the best work distribution across different processing units (CPU, GPU etc). It models static programming features such as the number of floatingpoint operations, and trains a SVM predictor with a set of programs, then statically chooses the best load-balancing solution.

Task-based programming and runtime systems for dynamic task scheduling are growing in popularity. While StarPU (see above) has been designed for heterogeneous systems from the beginning, support for GPUs has recently also been added to StarSS/OmpSS [4].

III. MULTI-VARIANT SKELETONS

*Skeletons* [9] are pre-defined generic software components derived from higher-order functions such as map, farm, scan and reduce, which can be parameterized in problem-specific user code and which implement certain frequently occurring patterns of control and data flow for which efficient target-specific implementations may exist.

*SkePU*[1] [19] is a C++ template library for portable programming of GPU-based systems that provides a simple and unified programming interface for specifying data-parallel and task-parallel computations with the help of pre-defined skeletons. All non-scalar operands of SkePU skeleton calls are passed in smart containers (see Section IV).

The SkePU skeletons provide multiple implementations, including CUDA and OpenCL implementations for GPU and multi-GPU execution as well as sequential and OpenMP implementations for CPU execution. The performance of these different implementations depends on the computation, system architecture and on the runtime execution context such as resource availability, problem size(s) and data locality. Such information can be generated off-line for each occurring *skeleton instance* (i.e., combination of a skeleton and parameterizing user function(s)) by a machine-learning approach [12], [17] and stored in a so-called *execution plan*. An execution plan is a run-time data structure attached to the skeleton instance for fast lookup of the expected best back-end (and possibly of recommended values for further, platform-specific tunable parameters such as the expected best number of threads or thread blocks to be used), given the current execution context of a call of the skeleton instance. By recomputing execution plans once for each new target system, SkePU can, at least to some degree, also achieve performance portability across a wide range of GPU-based multicore systems.

Besides this stand-alone scenario where SkePU programs can be executed directly (i.e., without a runtime system), there also exists a StarPU back-end for SkePU [15] where calls to data-parallel skeletons (map, mapoverlap, scan, reduce, ...) are, by partitioning of their operand containers, broken up into multiple independent or partly dependent tasks that are registered and scheduled dynamically in the StarPU runtime system. Each created task has again multiple implementations (OpenMP, CUDA, ...) that StarPU then can select from at runtime, using its own internal performance models. Note that the partitioning strategy and resulting subtask dependence structure for each skeleton is given by the skeleton's well-known computation pattern. This provides an increased level of task parallelism to StarPU (benefiting its dynamic scheduling and selection) and also allows hybrid execution on both CPU cores and GPUs.

Hybrid execution in SkePU [25] enables us to use multiple (types of) computing resources present in the system simultaneously for a given computation by dividing the work between them. Figure 2 shows the performance of executing a Coulombic potential application [38] written using SkePU skeletons on two GPU based systems. The computation can be executed across multiple GPU or CPU devices available

[1]SkePU is available at http://www.ida.liu.se/~chrke/skepu

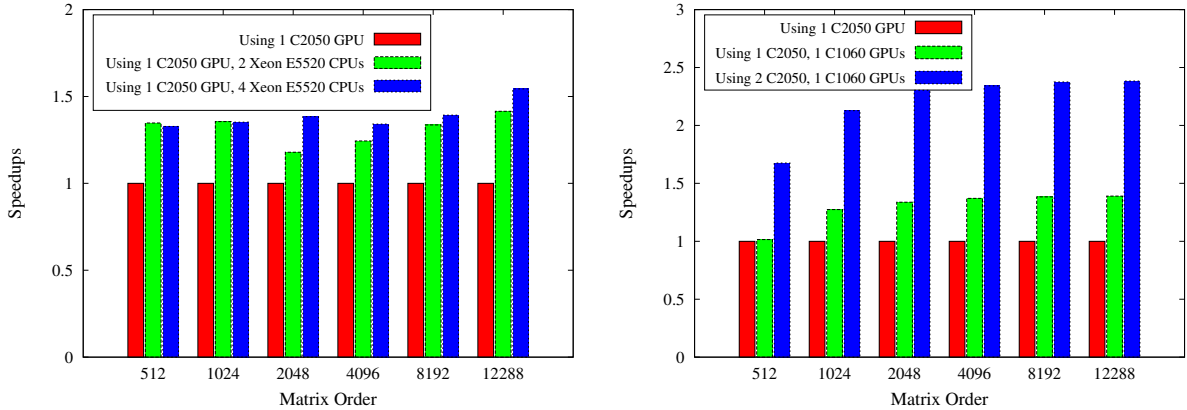

Figure 2. Coulombic potential grid execution on two GPU-based systems for different matrix sizes. The base-line is execution on a single GPU.

in the target system in a transparent manner.

Ongoing work on SkePU includes support for MPI, i.e. execution of SkePU programs over multiple nodes of a (GPU) cluster without any source code modification [31].

*Related Work*

*SkelCL* [34] is an OpenCL-based skeleton library that supports several data-parallel skeletons on vector and matrix container operands. The SkelCL vector and matrix containers provide memory management capabilities like SkePU containers. However, unlike SkePU, the data distribution for containers is exposed to the programmer in SkelCL.

The *Muesli* skeleton library, originally designed for MPI/OpenMP execution [8], has been recently ported for GPU execution [20]. It currently has a limited set of data-parallel skeletons which makes it difficult to port applications such as N-body simulation and Conjugate Gradient solver. Moreover, it lacks support for OpenCL and performance-aware implementation selection.

*Marrow* [32] is a skeleton programming framework for systems containing a single GPU using OpenCL. It provides data (map) and task parallel (stream, pipeline) skeletons that can be composed together to model complex computations. However, it focuses on GPU execution only (i.e., no execution on multicore CPUs) and exposes concurrency and synchronization issues to the programmer.

Support for GPU execution has recently been added [22] in the *FastFlow* library for pipeline and farm skeletons using OpenCL.

## IV. SMART CONTAINERS

In both SkePU and the PEPPHER composition tool, operands are stored and passed to component/skeleton calls in generic run-time container data structures. The container interfaces are designed to match existing C++ STL container types such as Vector and additionally encapsulate internal book-keeping information about the run-time state of the data, e.g. in which memory units, and where, valid copies of the container's elements can be found, and all element access is mediated through the containers by suitable operator overloading. Beyond element lookup, the containers provide the following services:

- Memory management
- Data dependence tracking and synchronization
- Communication optimization.

We refer to containers with such extended services performing automatic optimizations as *smart containers*.

A simple form of dynamic (greedy) communication optimization, provided by the first-generation smart containers in SkePU [19] and PEPPHER [16], is *lazy memory copying* to eliminate some redundant data transfers over the PCIe bus. This is particularly important in GPU-based systems as communication cost can be significant.

Essentially, each smart container internally implements a software memory coherence protocol for its payload data, enabling that existing valid copies closer to the executing unit can be reused but guaranteeing that stale copies are never accessed; if necessary, the currently valid (closest) copy holder is located and data transfer operations are triggered automatically to create a new copy in the device memory where it needs to be accessed. As partial copies might, at a given point of time, be spread over multiple memory units in the system, multiple data transfers from different sources might be required when accessing a given range of elements. The details can be found in [11, Ch. 4]. The communication optimization by SkePU smart containers can provide great speedups especially for applications doing many component/skeleton invocations, e.g. in a loop. The overhead of smart containers is negligible (less than 1%).

*Related Work*

NVIDIA *Thrust* [24] is a C++ template library that provides algorithms (reduction, sorting etc.) with an interface similar to C++ standard template library (STL). For vector data, it has the notion of device_vector and host_vector modeling data on host and device memory respectively; data can be transferred between two vector types using a simple vector assignment (e.g., `v0 = v1;`). Likewise, StarPU [3] provides smart containers at the granularity of full operands, with a partitioning mechanism to create subtask operands.

Note that smart containers could be extended to internally adapt not only the storage location but also the storage format (e.g., transposing a matrix, converting a sparse matrix, converting struct of arrays to array of structs or vice versa) and thereby become multi-variant, too. A case study of tuning variant selection with automated conversion between sparse and dense storage format for matrix operands has been considered in [1].

## V. GLOBAL COORDINATION AND COMPOSITION

Using smart containers for identifying dependences and taking care of synchronization, PEPPHER components and SkePU skeletons can be invoked asynchronously, generating task-level parallelism that can be exploited by the StarPU runtime system.

StarPU, PEPPHER and SkePU perform greedy, local selection among applicable variants at each call for the current call context. Greedy scheduling heuristics such as *Heterogeneous Earliest Finishing Time* (HEFT) [35] as provided e.g. in StarPU are popular and work well especially for independent tasks, but may lead, in programs with multiple (dependent) calls, to suboptimal overall performance as future costs of data transfers and selections affected by a greedy decision are not taken into consideration [13]. The *Global Composition Framework* [13], [11] takes a more global scope: for frequent dependence patterns such as chains and loops of dependent calls (as detected by static dependence analysis), it changes from HEFT to selection heuristics for multiple calls, such as *bulk selection*.

Another example of global coordination of multiple component calls is the task-parallel *pipeline pattern* implemented in the *PEPPHER transformation tool* [5] where a `while` loop can be pragma-annotated in the source code as a linear pipeline and blocks of statements in its body as pipeline stages; from these annotations the tool generates StarPU-task code and synchronization code for pipelined execution.

## VI. CONCLUSION

The PEPPHER component model with its non-intrusive XML annotations encourages an incremental way of porting C/C++ legacy codes to GPU-based systems, by componentizing the most frequently executed kernels of the application first and adding more implementation variants over time—the PEPPHER framework will detect automatically if a variant will be useful in a given runtime context.

SkePU skeletons are more convenient to use as all annotations and implementations are implicitly predefined, but where a computation cannot be expressed by the given set of skeletons it must still be parallelized by hand, maybe within a PEPPHER component.

GCF global composition is the most powerful approach because it has a larger scope of optimization than the greedy, call-local selection decisions made in SkePU, PEPPHER or StarPU, and it can benefit from source-code analysis and transformation but requires source code to be available and faces a much larger problem complexity [11].

*Acknowledgments* Research partly funded by EU FP7 projects PEPPHER (www.peppher.eu) and EXCESS (www.excess-project.eu) and by SeRC project OpCoReS.